\documentclass[%
 aip,
 amsmath,amssymb,
 reprint,%
]{revtex4-1}

\usepackage{graphicx}
\usepackage{dcolumn}
\usepackage{bm}
\usepackage{multirow}
\usepackage[utf8]{inputenc}
\usepackage[T1]{fontenc}
\usepackage{mathptmx}
\usepackage{etoolbox}

\usepackage[bookmarks=false]{hyperref}
\usepackage{xcolor}
\hypersetup{colorlinks=true,citecolor=blue,
linkcolor=blue,urlcolor=blue,pdfstartview=FitH,
bookmarksopen=true}

\newcommand{\ot}{\otimes}	
\newcommand{\kethjh}[1]{|{#1}\rangle}

\newcommand{\bkt}[2]{\langle{#1}|{#2}\rangle}
\newcommand{\ket}[1]{|{#1}\rangle}
\newcommand{\bra}[1]{\langle{#1}|}

\def\red#1{\textcolor{black}{#1}}

\makeatletter
\def\@email#1#2{%
 \endgroup
 \patchcmd{\titleblock@produce}
  {\frontmatter@RRAPformat}
  {\frontmatter@RRAPformat{\produce@RRAP{*#1\href{mailto:#2}{#2}}}\frontmatter@RRAPformat}
  {}{}
}%
\makeatother
\begin{document}

\preprint{AIP/123-QED}

\title{\red{Overcoming Residual Timing Jitter in Pump-Probe Interferometry via Weak Value Amplification and Deep Learning}}

\author{Jing-Hui Huang}
\author{Xiang-Yun Hu$^{*}$}%
 \email{xyhu@cug.edu.cn}

\affiliation{%
Hubei Subsurface Multiscale Image Key Laboratory, School of Geophysics and Geomatics, China University of Geosciences, Lumo Road 388, 430074 Wuhan, China.
}%

\date{\today}

\begin{abstract}
\red{
We introduce a hybrid methodology that synergistically combines weak value amplification (WVA) and deep learning to suppress the limiting effects of residual timing jitter in pump-probe interferometry, achieved through simulations of pump-induced time delays at a few-attosecond resolution. The WVA protocol, employing real weak values, amplifies the minute delay induced by sample perturbation, thereby translating it into a measurable shift of interference fringes.  However, this amplification introduces significant fringe distortion. To address this, we deploy deep learning architectures as high-precision parameter estimators: a convolutional neural network regressor (CNN-Regressor) for direct delay estimation and a classifier (CNN-Classifier) for discrete delay categorization. These are systematically benchmarked against traditional Fourier-transform-based analysis. Two key conclusions are drawn: (i) The WVA technique consistently enhances measurement precision across all estimators by effectively increasing the signal-to-noise ratio (SNR). (ii) Both deep learning models surpass the traditional FFT approach; the CNN-Regressor achieves a higher SNR at small weak values, while the CNN-Classifier enables accurate estimation even under a challenging "$\pi$ phase shift" condition where conventional analysis fails. This synergistic combination of WVA and deep learning establishes a powerful framework for attosecond metrology, paving the way for enhanced precision in ultrafast spectroscopy.
}
\end{abstract}


\maketitle 


\section{Introduction}
\red{
Pump-probe interferometry stands as a cornerstone technique in ultrafast spectroscopy, spanning applications from traditional femtosecond (fs) studies of electronic excitations in semiconductors~\cite{PhysRevLett.94.157402}, electron-phonon coupling in superconductors~\cite{PhysRevLett.64.2172} and 
photoinduced refractive index changes in transient absorption spectroscopy~\cite{tokunaga1992frequency,Tanghe2022,Tamming2019}
to modern attosecond (as) science. The advent of attosecond laser pulses has further revolutionized this field, enabling direct observation of electron dynamics in atoms~\cite{doi:10.1126/science.1189401}, molecules~\cite{doi:10.1126/science.1143679}, and solids~\cite{Cavalieri2007} through techniques such as attosecond streak cameras~\cite{Kienberger2004} and reconstruction of attosecond beating by interference of two-photon transitions~\cite{doi:10.1126/science.1059413}.
}

\red{
Despite these advances, a fundamental challenge persists: achieving the ultimate temporal resolution demanded by attosecond or femtosecond pulses requires controlling the pump-probe delay with attosecond precision~\cite{Vaughan:19}. In practice, this is limited by residual timing jitter between the pump and probe pulses, arising from beam pointing instabilities, thermal fluctuations, and fluctuating environmental conditions. Several active stabilization systems~\cite{10.1063/1.3475689,Chini:09,Mashiko:10,10.1063/1.4937623,Vaughan:19} have already been proposed to improve the stability of the timing jitter, especially for setting up a separate path for the pump and probe pulses. To the best of our knowledge, the residual timing jitter ($\delta t_{jitter}$) can be reduced to several tens of attoseconds root mean square error (RMS)~\cite{Vaughan:19}. Consequently, extracting few-attosecond pump-induced delays ($\tau$) under such noise conditions ($\tau< \delta t_{jitter}$) presents a significant metrological challenge.
}

Weak-value amplification (WVA) has attracted considerable interest in probing several fundamental tiny quantities~\cite{AAV,RevModPhys.86.307,Lundeen2011,PhysRevLett.111.033604,PhysRevLett.112.200401,10.1063/5.0164781,AWVA,PhysRevLett.125.080501,Xia2023,10.1063/5.0208376} \red{as well as quantitative phase imaging~\cite{PhysRevA.106.023518,https://doi.org/10.1002/lpor.202400092}}. The concept of ``weak value'' arises from performing a weak measurement on a pre-selected and post-selected system, resulting in an amplified value for the observable when the two states of the system are nearly orthogonal. 
Typically complex, the weak value $A_w$ relates to pointer shifts: the real part to position~\cite{PhysRevLett.134.080802}, the imaginary part to spectrum~\cite{PhysRevLett.111.033604}.
Recent work has demonstrated that WVA with real weak values can enhance phase sensitivity in interferometers~\cite{PhysRevLett.134.080802}, suggesting its potential for amplifying minute temporal delays in pump-probe configurations.

\begin{figure*}[t]
\centerline{\includegraphics[scale=0.07,angle=0]{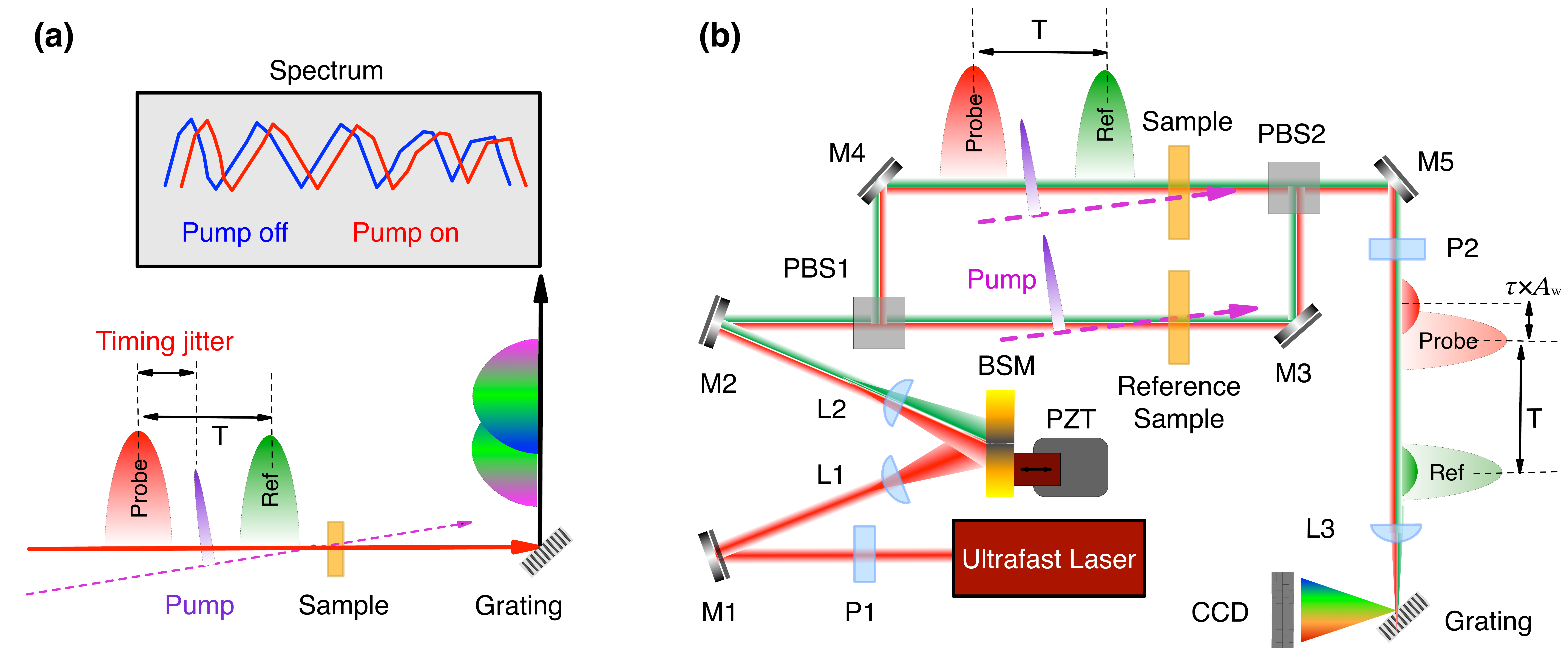}}
\caption{\label{Fig:Schemes_model}
\red{
\textbf{Schematic diagrams of pump-probe interferometry configurations.}
(a) Traditional pump-probe interferometry setup. The diagrams are adapted from Ref.~\cite{Tanghe2022}.
(b) Weak value amplification (WVA)-enhanced pump-probe interferometry implementation. 
The reference sample is subjected to the same pump pulse with identical timing jitter $\delta t_{\text{jitter}}$ as the main sample path, ensuring selective amplification of the sample-induced time delay $\tau$ while rejecting timing jitter (see main text for details). 
Key components: Ultrafast Laser (generating Fourier-transform-limited Gaussian pulses); P1, P2 (polarizers); M1-M5 (mirrors); L1-L3 (lenses); PBS1, PBS2 (polarization beam splitters); Grating (diffraction grating); CCD ( CCD camera).
 }
 }
\end{figure*}

\red{
Concurrently, deep learning has demonstrated remarkable capabilities in parameter estimation from noisy data across various domains~\cite{LeCun2015-wh,Rivenson:17,Rivenson2018-ts,Wiecha:21,Zuo2022}, including quantum metrology~\cite{Genty2021,Nolan2021,Ban_2021,Xiao2022,Rinaldi_2024} as well as quantum weak measurements~\cite{Liu_2019,Yang:24,Li:24}. For example, Liu \textit{et al}. proposed a weak measurement scheme with the assistance of a machine learning algorithm, and demonstrated a 6 dB reduction of mean-square error compared to the setup without machine learning~\cite{Liu_2019}. Yang \textit{et al}. theoretically and experimentally quantified entanglement of a class of two-photon polarization-entangled mixed states via weak measurements and deep learning technology~\cite{Yang:24}. Convolutional neural networks (CNNs) are particularly well-suited for analyzing 2D interferometric data, as their architecture naturally leverages spatial relationships and hierarchical patterns in image data. Therefore, data-driven deep learning has the potential to enhance the precision of WVA-based pump-probe interferometry by detecting 2D interferograms.
}

\begin{figure*}[t]
\centerline{\includegraphics[scale=0.24,angle=0]{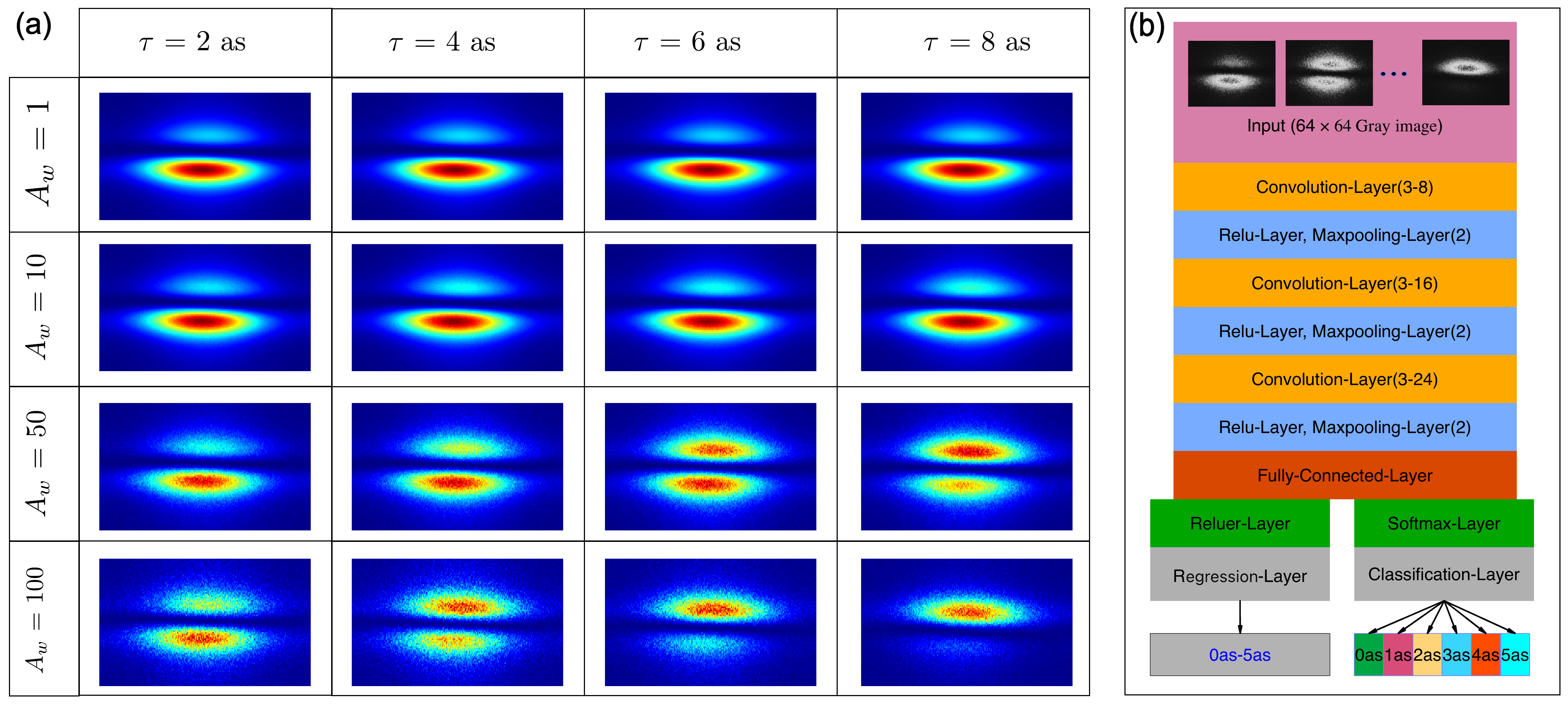}}
\caption{\label{Fig:different_tau}
\red{
\textbf{Simulated interferograms and deep learning architecture.}
(a) Spatial-spectral interferograms for pump-induced delays $\tau$ (0 as $< \tau <$ 10 as) at varying weak values $A_w$. Horizontal pixels correspond to spectral information (photon energy), while vertical pixels encode spatial information along the $y$-direction.
(b) Deep convolutional neural network architecture employed for parameter estimation in WVA-enhanced pump-probe interferometry.
}
}
\end{figure*}

\red{Here, we propose and numerically demonstrate a hybrid approach that combines WVA with deep learning to overcome residual timing jitter in pump-probe interferometry. As illustrated in Fig.~\ref{Fig:Schemes_model},
WVA enables the amplification of the pump-induced delay $A_w \tau$ using real weak values $A_w $, albeit at the expense of fewer detected photons [see Fig.~\ref{Fig:different_tau}(a)].
Then, two deep learning approaches with a final regressor layer or a classifier layer [see Fig.~\ref{Fig:different_tau}(b)] are studied to extract the delay parameter from the resulting 2D spatial-spectral interferograms. 
Through comprehensive numerical simulations across the attosecond regime (0 as < $\tau$ < 10 as) under realistic residual timing jitter conditions ($\delta t_{\text{jitter}}$ = 10 as RMS), we demonstrate that this synergistic approach achieves precision approaching the classical Fisher information (CFI) bound as $\tau$ increases~\cite{PhysRevLett.115.120401,PhysRevX.4.011032,PhysRevX.4.011031,PhysRevA.106.022619}. Our method robustly outperforms traditional Fourier-transform analysis~\cite{Guizar-Sicairos:08}, particularly in challenging operational regimes such as the $\pi$ phase shift condition where conventional methods fail due to fundamental limitations in phase unwrapping and linearity assumptions.
}

\section{WVA-Enhanced pump-probe interferometry}
\red{
The WVA protocol employing real weak values provides a general framework for enhancing optical interferometers sensitive to optical path differences (OPD)~\cite{PhysRevLett.134.080802}. In this work, we implement WVA within a modified pump-probe interferometric configuration~\cite{tokunaga1992frequency,Tanghe2022,Tamming2019} that incorporates a reference pulse for enhanced sensitivity. As illustrated in Fig.~\ref{Fig:Schemes_model}(a), this approach enables interference between probe and reference pulses following sample interaction, facilitating shot-to-shot normalization and direct measurement of transient phase shifts arising from pump-induced refractive index modifications.
}

\red{Unlike the previous WVA implementation employing two postselections in a double-slit interferometer~\cite{PhysRevLett.134.080802}, our scheme utilizes two distinct weak interactions within the pump-probe interferometer to detect pump-induced refractive index changes. These subtle refractive index variations manifest as both OPD modifications and corresponding temporal delays, which form the basis for our measurement protocol.
}

\red{Figure~\ref{Fig:Schemes_model}(b) schematically illustrates the integration of WVA into spatially resolved pump-probe interferometry for attosecond time-delay measurements. 
The experimental configuration employs active stabilization systems based on Michelson interferometers to control timing delays and suppress residual jitter~\cite{10.1063/1.3475689,Chini:09,Mashiko:10,10.1063/1.4937623,Vaughan:19} (not shown in schematic). Probe and reference pulses are generated by splitting a single pulse via a reflection beam-splitting mirror (BSM)~\cite{Tzallas2003,PhysRevLett.102.213904,doi:10.1073/pnas.1407421111,Mashiko:20}, ensuring co-propagation along a common optical path through the sample and spectrometer. This configuration enables precise control of the temporal separation $T$ between pulses via the BSM displacement. 
The reference pulse temporally precedes the probe pulse, with the pump pulse positioned between them. Consequently, the timing jitter $\delta t_{\text{jitter}}$ between the pump and probe pulses introduces temporal fluctuations into the temporal separation $T$, resulting in an effective temporal interval of $T + \delta t_{\text{jitter}}$ between the probe and reference pulses.
Crucially, the pump-induced delay $\tau$ is introduced in only one arm of the polarized Mach-Zehnder interferometer, consistent with established WVA methodologies for measuring birefringence-induced delays~\cite{PhysRevLett.111.033604,PhysRevLett.134.080802}. This asymmetric implementation enables amplification of $\tau$ through the standard WVA protocol.
}

\red{
In the optical configuration depicted in Fig.~\ref{Fig:Schemes_model}(b), a reference sample is subjected to the same pump pulse, experiencing identical timing jitter $\delta t_{\text{jitter}}$ as the main sample path. This design offers two crucial advantages. First, the reference module operates at the point of maximum sensitivity for detecting the pump-induced time delay $\tau$. More importantly, employing the same pump source for both interferometer arms ensures that any timing jitter $\delta t_{\text{jitter}}$ manifests as a common-mode disturbance, thereby preventing the introduction of additional optical path differences in the polarized Mach-Zehnder interferometer for the probe. Without this common-path design, the timing jitter between the two probe paths would manifest as apparent birefringence delays that would also undergo amplification through the WVA protocol. Consequently, this optical arrangement ensures that the WVA-enhanced pump-probe interferometry selectively amplifies the sample-induced time delay $\tau$ while rejecting timing jitter $\delta t_{\text{jitter}}$, thereby realizing the fundamental WVA advantage quantified in Equation~(\ref{Eq:final-selection-pointer}).
}

In standard WVA protocol, WVA involves four processes: initially preparing the system at state $\kethjh{\Phi_i}$ and the two pointers at states $\kethjh{\Psi_{i,1}}$ and $\kethjh{\Psi_{i,2}}$; weak interaction between the system and the pointer by the unitary operator $\Hat{B}= {\rm exp}(-ic\tau \hat{p}\ot \hat{A})\,$ with a observable operator $\hat{A}=\ket{H} \bra{H}-\ket{V} \bra{V}$ and moumentum operator $\hat{p}$; post-selection of the system at state $\kethjh{\Phi_f}$; estimation of coupling strength $c\tau$ from the spatial-spectral interferograms caused by the interference between the final pointer states $\kethjh{\Psi_{f,1}}$ and $\kethjh{\Psi_{f,2}}$. The bases $\kethjh{H}$ and $\kethjh{V}$ represent the horizontal and vertical polarized states of the photons, respectively.

The ultrafast laser generates a temporal pointer with a Gaussian temporal envelope $f(t)$ and transverse spatial distribution $\widetilde{U}(x,y)$, where $(x,y)$ are beam cross-section coordinates.
The two pointers, located at the $(x_1,y_1)$ coordinate on the beam-splitting mirror (BSM), can be defined by:
\begin{eqnarray}
\label{Eq:pre-selection-pointer}
{\bkt{t}{\Psi_{i,1}}}(x_1,y_1;t) 
&=&E_0 e^{- {t^2}/{ \sigma^2} }  e^{ - {(x_1^2+y_1^2)}/{ \omega_1^2} } e^{i \Omega t}, \\
{\bkt{t}{\Psi_{i,2}}}(x_1,y_1;t) 
&=&E_0 e^{- {(t-T)^2}/{ \sigma^2} }  e^{ - {(x_1^2+y_1^2)}/{ \omega_1^2} } e^{i \Omega t},
\end{eqnarray}
where the delay $T$ represents the temporal distance between the probe and reference in pump-probe interferometry~\cite{Tokunaga:92}, $\sigma$ and $\omega_1$ are the pulse width and transverse beam diameter, respectively. The angular frequency of the photon is denoted by $\Omega$.  The two pointers in WVA also act as the probe and reference beams in pump-probe interferometry, respectively. 
The width $\sigma$ is selected to ensure $\sigma \gg \tau$, thereby maintaining the weak measurement regime required for WVA.
Polarizer P1 is used to pre-select the system at state $|{\Phi_i}\rangle=(|{H}\rangle + |{V}\rangle)/ \sqrt{2}$, while polarizer P2 post-selects the system at state $\kethjh{\Phi_f}=\cos{(\beta - \pi/4)} \ket{H} + \sin{(\beta - \pi/4)}\ket{V}$.  
Here, $\beta$ represents the angle between the polarization direction of P2 and the orthogonal polarization direction of P1. When the states $\kethjh{\Phi_i}$ and $\kethjh{\Phi_f}$ of the system are nearly orthogonal, the weak value can be calculated by
\begin{eqnarray}
\label{Eq:Aw}
A_{w}:=\frac{\bra{\Phi_{^{f}}}\hat{A}\ket{\Phi_{i}}}{\bkt{\Phi_{^{f}}}{\Phi_{i}}},
\end{eqnarray}
which can significantly exceed the eigenvalue and result in an amplified temporal shift $A_w \tau$, at the cost that the probability $p=|{\bkt{\Phi_{^{f}}}{\Phi_{i}}}|^2 \simeq 1/{A_w}^2$ of successful post-selection decreases~\cite{PhysRevLett.105.010405}.
Unlike standard WVA, our approach requires detecting spatial-spectral interferograms of the synthesized electric field at the focal point $(x_2, y_2)$ on the grating. This synthesized electric field, denoted as $\widetilde{E}=\widetilde{E}(x_2, y_2; t)$, can be calculated from the superposition of the final two pointers:
\begin{equation}
\begin{aligned}
\label{Eq:final-selection-pointer}
\widetilde{E}&={\bkt{t}{\Psi_{f,1}}}(x_2,y_2;t) + {\bkt{t}{\Psi_{f,2}}}(x_2,y_2;t)  \\
&  =  \frac{1}{A_w} \bigg ( f(t) \big | \widetilde{U}(x_2,y_2) \big | \times {\rm e}^{i \varphi_1(x_2,y_2)}  \\
&+  f(t-T-\delta t_{jitter}- A_w \tau) \times\big | \widetilde{U}(x_2,y_2)\big | \\
& \times {\rm e}^{i \varphi_2(x_2,y_2)} e^{i\Omega(T+\delta t_{jitter}+A_w \tau)} \bigg), 
\end{aligned}
\end{equation}
with the profile $\widetilde{U}(x_2,y_2)$ and the spatial phase $\varphi(x_2,y_2)$ of the focused beam defined by the diffraction theory~\cite{Mashiko2003}.
The relative phase $\varphi_r=\Omega(T+A_w \tau)$ depends on the overall time delay. 
Although the two split beams, i.e., the final pointers in WVA, have the same electric-field profiles $\left | \widetilde{U}_{}(x_2,y_2) \right | $ at the focal point, the wavefronts $\varphi_{1}(x_2,y_2)$ and $\varphi_{2}(x_2,y_2)$ tilt in opposite directions to each other, resulting in interference.
Finally, the spatial-spectral interferograms  $N_{p}(x_2,y_2)$ dependent on $\tau$ can be obtained from the superimposed electric field $\widetilde{E}(x_2,y_2;t)$ by performing a Fourier transformation with respect to time $t$.

\begin{table}[b]
\caption{\label{Tab_parameters}
\red{
\textbf{Simulation parameters for WVA-enhanced pump-probe Spectroscopy.} 
{The simulation includes the WVA setup, the laser sources, the residual timing jitter $\delta t_{jitter}$, and the detection noise.}
 }
 }
 \renewcommand{\arraystretch}{1.2}
\begin{tabular}{l@{\hspace{0.65cm}}l@{\hspace{0.65cm}}l@{\hspace{0.65cm}}l}
\toprule
                         Type  & Parameter        & Meaning                      & Value                              \\ \hline
\multirow{4}{*}{\rotatebox[origin=c]{90}{WVA}}      
& $\tau$         & Time delay          & $1-10$ $\rm as$                           \\
 & $\delta t_{jitter}$ & Timing jitter                   & $10$ as RMS    \\
 & ${\rm Re} A_{w}$ & Weak value                   & $1-100$     \\
& $\lambda$& Wavelength                  & 800 nm  \\ \hline
\multirow{5}{*}{\rotatebox[origin=c]{90}{Laser}}   
& $P^{0} $        & Pulse energe                    & $600 \, {\rm \mu J}$               \\
& $\lambda_0$      &  Central wavelength            & 800 nm             \\
& $\sigma$      & Pulse width             & 8 fs             \\
& $F_r$      & Repetition Rate             & 10 kHz             \\
                           & $T$              &  Temporal distance                 & $1000$ $\rm as$               \\
     \hline
\multirow{4}{*}{\rotatebox[origin=c]{90}{Detection}} & $\eta$          & Quantum efficiency              & 0.5           \\
& $D_p$            & Pixel size                  & $13.5$ $\mu$m\\
& $t_e$            & Exposure time               & 1 s\\
& $\mathcal{N}(0,\sigma_{r}^2)$            &  Read-out noise             & $\mathcal{N}(0, (2.5 \,e^{-})^2)$ \\
\toprule
\end{tabular}
\end{table}

\begin{figure*}[t]
\centering
\includegraphics[width=0.99\textwidth]{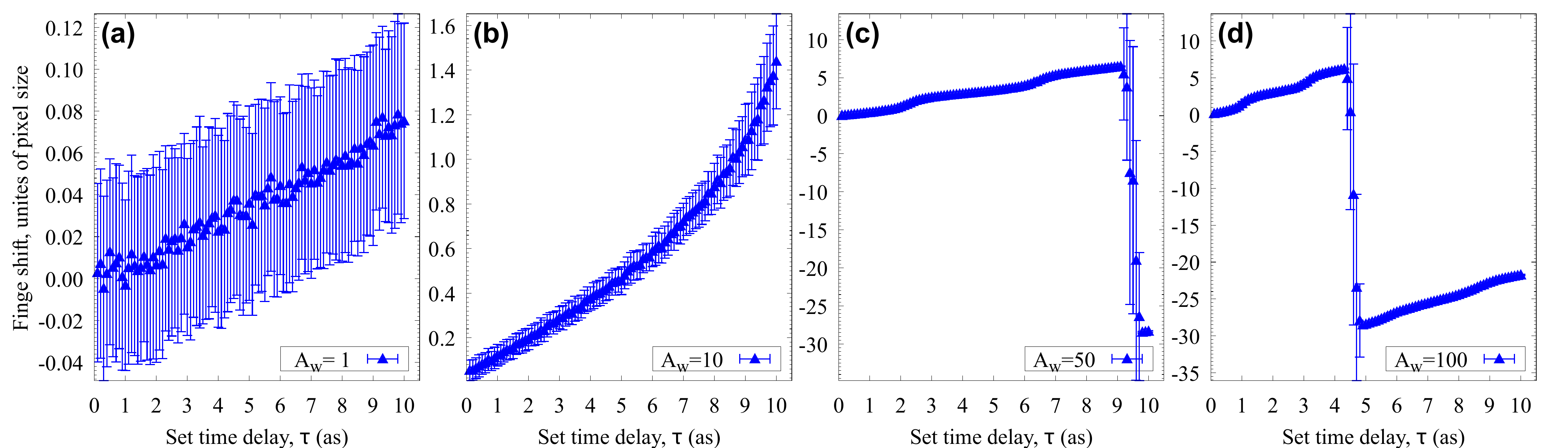}
\caption{\label{Fig:FFT_reslut} 
\red{
\textbf{Traditional FFT analysis of fringe shifts.}
Fringe shift measurements derived via FFT-based subpixel registration for pump-induced delays $\tau$ at varying weak values $A_w$. Note the emergence of $\pi$ phase shift discontinuities in panels (c) and (d), which violate the linear phase assumption of conventional interferometric analysis and introduce significant estimation errors.
}
}
\end{figure*}

\red{Equation~(\ref{Eq:final-selection-pointer}) encapsulates the principal advantage of the WVA protocol: the direct amplification of the pump-induced delay $\tau$ by the weak value $A_w$. This amplification effectively enhances the signal-to-noise ratio for the delay parameter, thereby improving the precision of estimating $\tau$ against the background of residual timing jitter $\delta t_{\text{jitter}}$. 
}

\red{
Note that the weak value $A_w$ in the denominator of Equation~(\ref{Eq:final-selection-pointer}) implies a post-selection probability scaling as $p \propto 1/|A_w|^2$, which reduces both the detected photon counts and the CFI calculated in Appendix B. In this work, the CFI calculation incorporates both shot noise and read-out noise while excluding residual timing jitter, thereby establishing the theoretical precision limit for time delay estimation. As shown in Fig.~\ref{Fig:CNNtraining}, the CFI bound remains constant for $A_w>1$ and equals the CFI at $A_w=1$, confirming the prediction by A. N. Jordan \textit{et al.} that WVA with real weak values concentrates the complete Fisher information in the post-selected events while maintaining optimal estimation precision despite the reduced photon counts~\cite{PhysRevX.4.011031}.
}

\section{Result of parameter estimation}
\red{
To systematically investigate the proposed WVA-enhanced pump-probe interferometry, we performed simulations of pump-induced delay ($\tau$) measurements across the range 0 as $< \tau <$ 10 as for varying weak values ($A_w$). The simulations incorporated realistic residual timing jitter conditions ($\delta t_{\text{jitter}}$ = 10 as RMS) and detection noise, including both shot noise and read-out noise. Key simulation parameters, including the WVA configuration, pulse characteristics (energy and width), and detection conditions, are summarized in Table~\ref{Tab_parameters}. 
The methodology for introducing shot noise and read-out noise to the ideal spatial-spectral interferograms $N_{p}(x_2,y_2)$ is detailed in Appendix A.
Simulated interferograms are available in the public dataset.
}

\red{Figure~\ref{Fig:different_tau}(a) presents spatial-spectral interferograms simulated across the range 0 as $< \tau <$ 10 as for varying weak values ($A_w$). The condition $A_w = 1$ corresponds to conventional spatially resolved pump-probe interferometry without WVA, where $\tau$-dependent variations in the interferograms are imperceptible to visual inspection. As $A_w$ increases, the amplification induces progressively more pronounced fringe shifts and shape modifications. However, this enhancement comes at the cost of reduced signal-to-noise ratio, manifesting as increased blurring in the interferometric patterns. This work systematically evaluates and compares the precision of time delay estimation using both traditional FFT analysis and deep learning approaches.
}

\subsection{Traditional FFT approach} 
\red{
The conventional method for estimating the time delay $\tau$ from 2D interferograms relies on detecting pure translational shifts between interferograms acquired at different $\tau$ values. In this work, we compute these fringe shifts using a widely adopted subpixel registration algorithm that employs an upsampled cross-correlation calculated via fast Fourier transform (FFT)~\cite{Guizar-Sicairos:08}.
}

\red{
Figure~\ref{Fig:FFT_reslut} presents the FFT-derived fringe shifts for measuring pump-induced delay $\tau$ across varying $A_w$. While statistical errors are expected to decrease with increasing weak values, a significant challenge emerges at higher amplification factors: the appearance of $\pi$ phase shifts, where fringe positions change dramatically near critical "$\pi$ phase points." In optical interferometry, such $\pi$ phase shifts represent a fundamental limitation as they violate the core assumption of linear phase relationships underlying conventional analysis.
Although the nonlinear relationships observed in Fig.~\ref{Fig:FFT_reslut}(c) and Fig.~\ref{Fig:FFT_reslut}(d) can be calibrated and potentially utilized for $\tau$ estimation, data in proximity to $\pi$ phase points introduce substantial errors that degrade measurement accuracy. The comparative SNR achieved by the traditional FFT approach is further analyzed in Fig.~\ref{Fig:CNNtraining}.
}

\subsection{Deep learning estimator} 

We train deep CNNs via supervised learning to capture nonlinear relationships between 2D interferograms and $\tau$.
Supervised learning necessitates a training dataset containing input examples (here, 2D spatial-spectral interferograms) paired with labeled target outputs (the corresponding $\tau$ values). 
As depicted in Fig.~\ref{Fig:different_tau}(b), two distinct architectures are implemented: a CNN-Regressor for regression and a CNN-Classifier for classification. Both architectures share identical deep CNN backbone structures but differ in their final layers: the regressor terminates with a linear layer optimized via root mean square error (RMSE) minimization, while the classifier concludes with a softmax-activated layer trained using cross-entropy loss.
The input to both networks is a fixed-size $64 \times 64$ grayscale interferogram. Feature extraction proceeds through a sequence of convolutional layers, each followed by a rectified linear unit (ReLU) activation and max-pooling operations. These layers feed into a fully connected layer, which connects to the task-specific final output layer.

The CNN-Regressor and CNN-Classifier models were trained under identical conditions: an initial learning rate of 0.001, a maximum of 1500 iterations, and a Training/Validation split of 8000:2000 samples. 
For the CNN-Regressor, each interferogram is assigned a continuous $\tau$ value as its label. 
In contrast, the CNN-Classifier categorizes $\tau$ into 100 discrete classes (0 as, 0.1 as 0.2 as... 9.9 as), where each class represents a bin centered at a specific $\tau$ value. To emulate realistic experimental variability, the actual $\tau$ values within each class are randomized uniformly around the bin center with a tolerance of $\pm$ 0.05 as, representing the time jitter.

\begin{figure*}[t]
\centering
\includegraphics[width=0.99\textwidth]{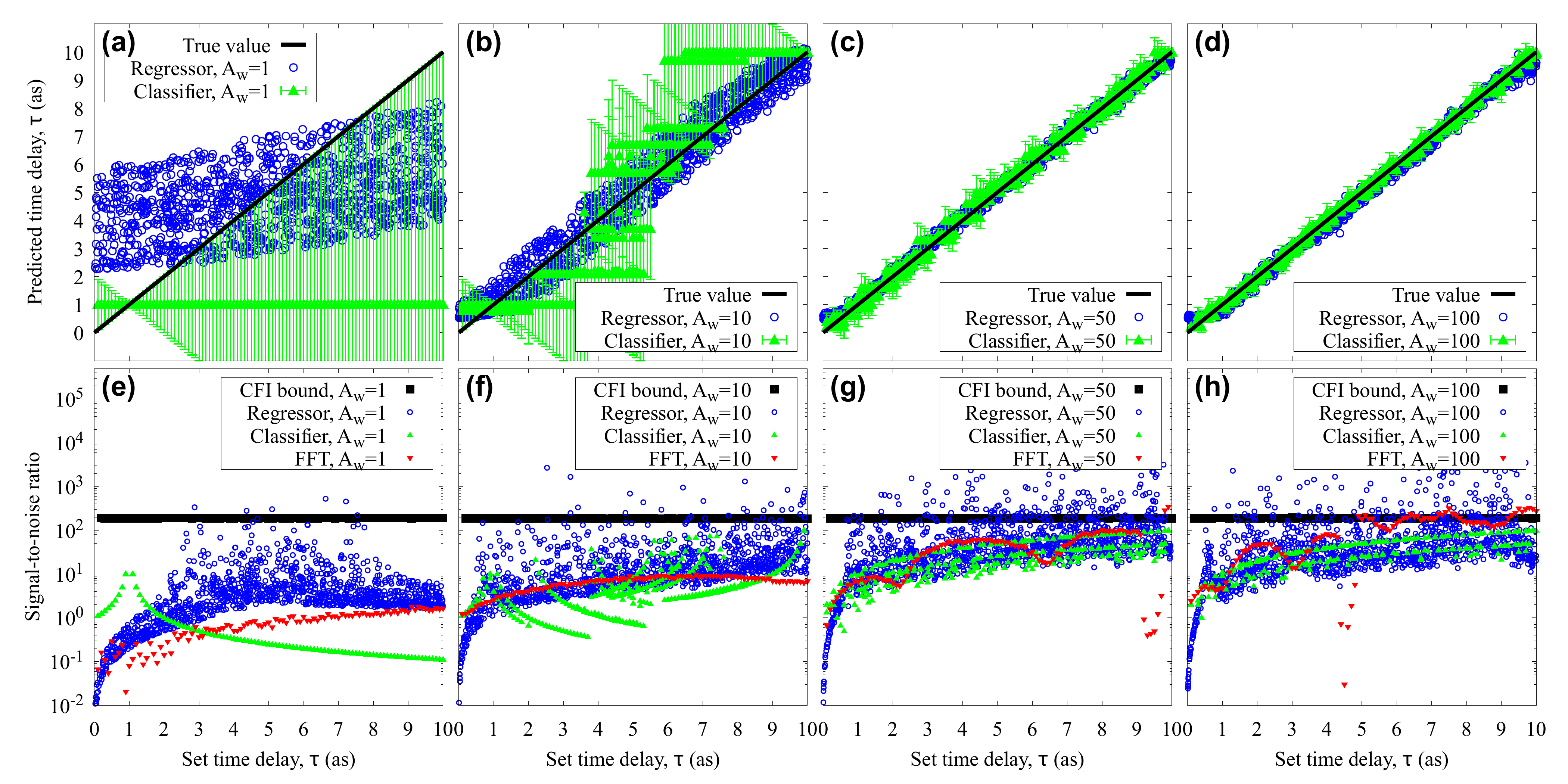}
\caption{\label{Fig:CNNtraining} 
\textbf{Deep learning performance for time delay estimation.}
(a)-(d) Comparison between predicted and actual time delays $\tau$ across varying $A_w$ values. (e)-(h) Signal-to-noise ratio performance relative to the classical Fisher information (CFI) bound and the traditional FFT approach. Each data point represents a single measurement instance.
\red{The SNR values for the traditional FFT approach were calculated from the data in Fig.~\ref{Fig:FFT_reslut} as the mean divided by the standard deviation of the fringe shift measurements.}
}
\end{figure*}

Figure~\ref{Fig:CNNtraining} compares the estimated time delays from both networks across varying weak values. 
We found that the performance of the deep learning model is strongly task-dependent, with the CNN-Regressor outperforming the CNN-Classifier at $A_w = 1$ [Figs.~\ref{Fig:CNNtraining}(a) and~\ref{Fig:CNNtraining}(e)], achieving attosecond precision with smaller deviations from the true values $\tau$ and higher SNRs.
However, as the weak value $A_w$ increases, the CNN-Classifier demonstrates enhanced accuracy across the entire delay range.

\red{
The data consistently demonstrate that WVA enhances measurement precision for both network types by effectively increasing the signal-to-noise ratio. At $A_w=50$ and $A_w=100$, the achieved SNRs have the potential to approach the CFI bound as $\tau$ increases. Notably, however, increasing the weak value from $A_w=50$ to $A_w=100$ does not yield further SNR improvement, as shown in Figs.~\ref{Fig:CNNtraining}(g) and~\ref{Fig:CNNtraining}(h). We attribute this saturation effect to the reduced photon detection rates resulting from the postselection process in WVA.
}

\section{Discusssion and Summary}

For real experimental implementation and dataset construction, classification substantially outperforms regression in WVA-based attosecond delay estimation due to its tolerance for timing jitters ($\pm$0.05 as bin tolerance vs. exact continuous labels), robustness to low-SNR interferograms, reduced calibration complexity (discrete bins vs. absolute delay precision), and resilience to experimental imperfections, enabling near-CFI-limited performance at high weak values where regression fails \red{at small time delays}, while simultaneously streamlining training data acquisition through practical bin-based labeling compatible with attosecond metrology constraints.

\red{
When comparing the performance between deep learning and the traditional FFT approach, the CNN-Regressor achieves higher SNR at small weak values, while the CNN-Classifier enables accurate estimation even under the challenging "$\pi$ phase shift" condition where conventional FFT analysis fails. These results agree with previous studies on machine-learning-assisted weak measurement~\cite{Liu_2019,Yang:24,Li:24}. Our findings indicate that deep learning can extract more comprehensive physical information from the interferograms, including complex morphological changes in fringe patterns, compared to the traditional FFT approach which relies solely on fringe shifts.
}

\red{
The few-attosecond timing resolution achieved in our protocol using femtosecond pulses corresponds to OPD measurements with few-nanometer path length resolution, enabling the detection of physical processes occurring on femtosecond timescales in spatially resolved spectral phase pump-probe measurements. Our choice of 8-fs pulses with 600-µJ energy for both pump and probe~\cite{Vaughan:19} is motivated by the practical constraint that current attosecond laser sources cannot provide sufficient photon flux for WVA postselection, which would otherwise result in detection-noise-limited performance.
}

\red{
The schematic in Fig.~\ref{Fig:Schemes_model}(b) can be further modified to incorporate extreme-ultraviolet attosecond pump pulses generated via high harmonic generation~\cite{Vaughan:19}. Using femtosecond probe and reference pulses in this configuration could enable the detection of attosecond-scale physics through averaging of the pump-induced time delay. A more direct approach would employ attosecond pulses for both pump and probe, where the increased fringe density in spatial-spectral interferograms~\cite{Mashiko:20} could enhance precision through machine-learning-assisted WVA. Therefore, the proposed machine-learning-assisted WVA framework could be extended to spatially resolved spectral phase interferometry using entirely attosecond sources when higher-energy pulses become available.
}

\red{
In summary, this work introduces a hybrid methodology that synergistically combines weak value amplification (WVA) and deep learning (DL) to overcome the limiting effects of residual timing jitter in pump-probe interferometry, enabling the measurement of pump-induced attosecond time delays. The WVA protocol, employing real weak values, amplifies minute temporal delays into measurable fringe shifts, while convolutional neural networks (CNNs) act as efficient parameter estimators from the resulting spatial-spectral interferograms. We demonstrate that this synergy enables precision approaching the classical Fisher information bound, surpassing traditional FFT analysis. Systematic comparison reveals a task-dependent performance: a CNN-regressor excels at lower weak values, whereas a CNN-classifier dominates under high amplification, proving robust even at the challenging $\pi$ phase shift condition where conventional methods fail. 
Our findings suggest that the combined WVA-DL framework could be extended to other ultrafast metrology tasks, such as measuring optical rotations, orbital angular momentum, or spatio-temporal field correlations, where weak measurements can amplify subtle signals and deep learning can robustly decode them from noisy, complex data.
}
\\

\begin{acknowledgments}
We are in debt to Jeff. S. Lundeen and Kyle M. Jordan at the University of Ottawa for stimulating discussions.
This study was supported by the National Natural Science Foundation of China (Grants No.~42327803 and No.~42504048). 
J-H. Huang acknowledges support from the Hubei Provincial Natural Science Foundation of China (Grant No.~20250650025), the Fellowship Program of China National Postdoctoral Program for Innovative Talents under Grant Number BX20250161.
\end{acknowledgments}

\section*{Data Availability Statement}

Data (simulated interferograms for the deep learning) openly available in a public repository~\cite{DVN/VQANT5_2025} that issues datasets with DOI: https://doi.org/10.7910/DVN/VQANT5

\appendix

\section{Simulation parameters}
\label{Sec:simulation_parameters}
\red{
The simulation parameters in Fig.~1 were selected in consideration of practical engineering limits, following the configuration of an attosecond pump-probe interferometer that uses 8-fs pulses with an energy of 600-µJ~\cite{Vaughan:19}. Specifically, parameters include a Fourier-transform-limited Gaussian pulse with a width of $\sigma$ = 8-fs and a center photon energy of $E_e$ = 1.55-eV, a temporal distance of $T$ = 1000-as, a spot size of $\omega_1$ = 600-$\rm \mu m$ at BSM, and a spot size of $\omega_2$ = 60-$\rm \mu m$ at the Grating. The initial pulse strength can be calculated by considering the pulse with $P_0$ pulse energy and 10-KHz repetition rate}:
\begin{equation}
\label{initial pulse strength}
\begin{aligned}
E_0 = \left (  \frac{2.0 \times P_0  \times  (2 / \pi)^{\frac{3}{2}} }{ \sigma \omega_1^2 c \varepsilon_0 } \right )^{\frac{1}{2}},
\end{aligned}
\end{equation}
where $\varepsilon_0=  8.854 \times 10^{-12}$-F/m is the vacuum dielectric factor.
In our study, we perform simulations to determine the number of photons $N_p(x_2, y_2)$. 
The number of photons incident on each pixel of PIXIS-NIR-CCD can be calculated by
\begin{equation}
\label{photon number}
\begin{aligned}
N_p(x_2, y_2) \approx \frac{\mathcal{F}^2(x_2,y_2) \times  \sigma  c \varepsilon_0 D_p^2}{ h f \times 1.62 \times 10^{-19} {\rm J/eV}} \times t_{e} \times 10000/s,
\end{aligned}
\end{equation}
where $hf/eV$ denotes the single photon energy at each pixel with a pixel size of $D_p$ = 13.5-$\mu$m.
The exposure time of the PIXIS-NIR-CCD is denoted as $t_e$ and is set at $t_e$ = 1-s. The value of 10000/s signifies 10000 pulses per second, corresponding to a repetition time of 10-kHz.
Considering the inherent noise variation representing shot noise, the actual CCD received photoelectronic $M_e(x_2, y_2)$ follows a Poisson distribution:
\begin{equation} \begin{split}
\label{Eq:Posson_distribution_simulation}
p(M_e(x_2, y_2))
&=\frac{ (\eta N_p(x_2, y_2))^{M_e(x_2, y_2) e^{-\eta N_p(x_2, y_2) } }}{ M_e(x_2, y_2) ! }.
\end{split} \end{equation}
\red{In our simulation, a deep-cooled, near-infrared optimized CCD camera (Teledyne Princeton Instruments PIXIS-NIR) was employed, featuring low readout noise of 2-3 $e^-$ RMS.
}
When the PIXIS-NIR-CCD receives a few photon counts, the read-out noise plays an important role and follows the additive Gaussian distribution~\cite{van1998image}:
$\mathcal{N}(\mu_{r},\sigma_{r}^2)$, in which $\mu_{r}=$ 0-$ e^-$ and $\sigma_{r}^2= $ $(2.5 \, e^-)^2$.
Finally, detector outcomes $N_e(x_2, y_2)$ considering the shot noise and read-out noise can be calculated as follows:
\begin{equation}
\label{photon number2}
\begin{aligned}
N_e(x_2, y_2) = M_e(x_2, y_2) + \mathcal{N}(\mu_{r},\sigma_{r}^2).
\end{aligned}
\end{equation}
Consequently, the spatial-spectral interferograms, exemplified in Fig.~2, illustrate the correlation between the number of electrons and the two pixels.

\section{Classical fisher information}
\label{Sec:CFI}
The classical fisher information bound in Fig.~3 represents the Cram{\'e}r-Rao bound, which gives a lower bound on the variance and a higher bound on the SNR of any unbiased estimator of the time delay. The bound  is directly related to the classical Fisher information (CFI) and CFI can be calculated by:
\begin{equation} 
\label{Eq:define_FisherInformation}
{F}(\tau)
 = \sum\limits_{m} \sum\limits_{n} p({N_e}(m,n) | \tau, X) 
 \times \left[  \frac{\partial}{\partial \tau} {{\rm ln} {\,} p({N_e}(m,n) | \tau, X)}  \right]^{2},
\end{equation}
where $m$ and $n$ are the pixel indices along $y$- and $x$- directions, respectively. The distribution ${N_e}(m,n)$ represents the X-CCD outcomes simulated based on the equation (\ref{photon number2}). Then, the CFI bound is calculated as
\begin{equation} 
\label{Eq:define_FisherInformation}
{\rm SNR}
 = \frac{\tau}{ \sqrt{1/ {F}(\tau)}}.
\end{equation}

\section{CNN Training}

\begin{figure}[htp!]
\centerline{\includegraphics[scale=0.32,angle=0]{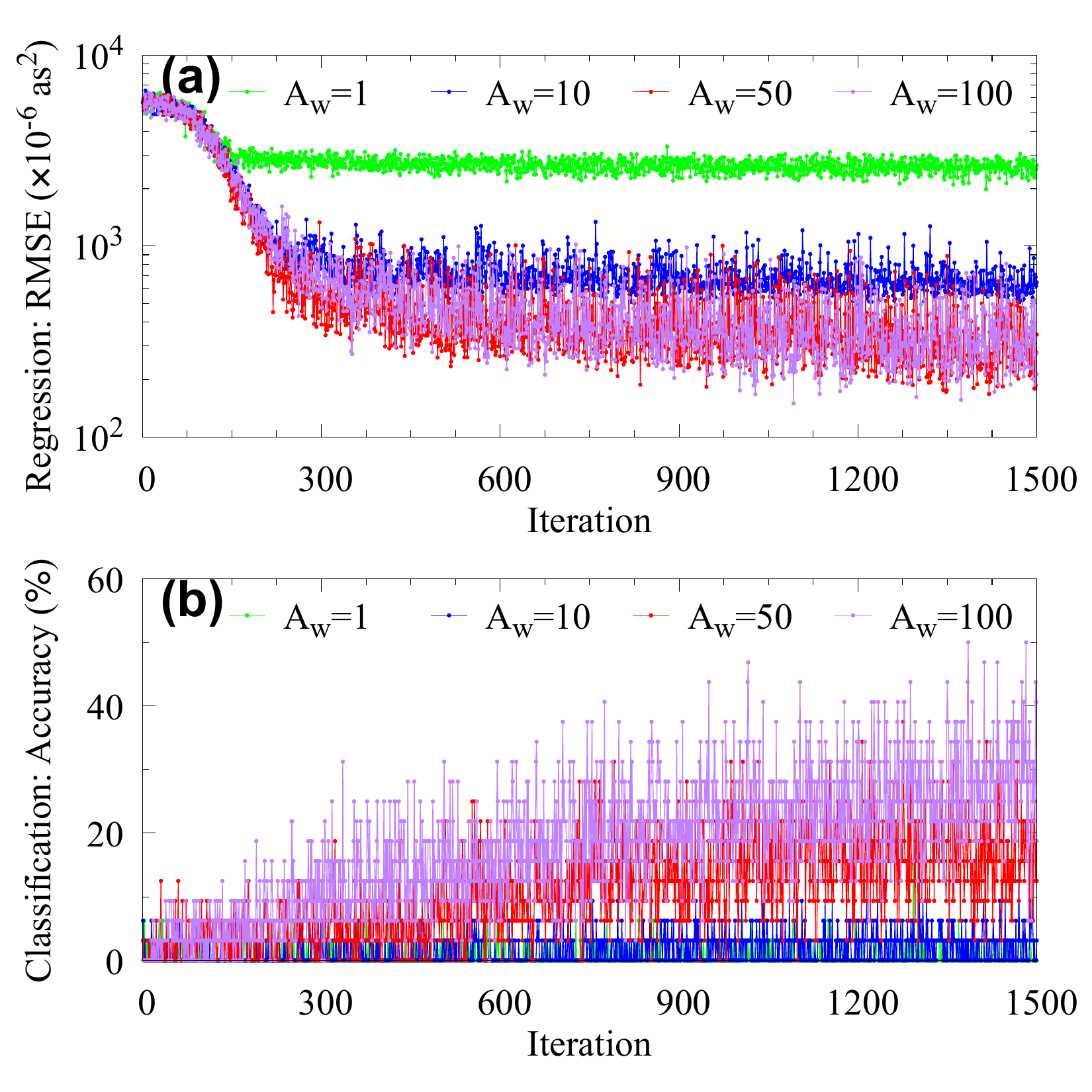}}
\caption{\label{Fig:CNNtraining_detail}
Estimator for the WVA-based pump-probe interferometry: deep convolutional neural networks.
(a) Root mean square error (RMSE) in various $A_w$ as a function of the iteration number within the CNN-Regressor network. (b) The accuracy of the training on various $A_w$ as a function of the number of iterations within the CNN classifier network.
 }
\end{figure}

The training processes of the CNN-Regressor and CNN-Classifier are shown in Fig.~\ref{Fig:CNNtraining_detail}(a) and in Fig.~\ref{Fig:CNNtraining_detail}(b), respectively.
For the CNN-Regressor, the root mean square error decreases monotonically with training iterations, converging at the maximum epoch limit.
For the CNN-Classifier, we verified that the training and validation curves overlap, indicating the absence of overfitting or underfitting in the CNN training results.
\\




\bibliography{aipsamp}

\end{document}